\documentclass[12pt,a4paper]{article}
\usepackage[english]{babel}
\usepackage{xcolor}
\usepackage{amsmath, amssymb, amsthm, mathtools} 
\usepackage{mathrsfs} 
\usepackage{bbm}

\usepackage{enumitem}
\setlist{nosep} 
\usepackage{hyperref}

\theoremstyle{plain}
\newtheorem{theorem}{Theorem}[section]
\newtheorem{lemma}[theorem]{Lemma}
\newtheorem{proposition}[theorem]{Proposition}
\newtheorem{corollary}[theorem]{Corollary}

\theoremstyle{definition}
\newtheorem{definition}[theorem]{Definition}
\newtheorem{example}[theorem]{Example}
\newtheorem{remark}[theorem]{Remark}

\newcommand{\R}{\mathbb{R}}
\newcommand{\Cm}{\mathbb{C}}
\newcommand{\N}{\mathbb{N}}

\newcommand{\E}{\mathbb{E}}
\newcommand{\Prb}{\mathbb{P}}
\newcommand{\Tr}{\operatorname{Tr}}
\newcommand{\Hilb}{\mathcal{H}}
\newcommand{\Sfr}{\mathfrak{S}}
\newcommand{\Mbf}{\mathbf{M}}
\newcommand{\Dom}{\mathcal{D}}
\newcommand{\opF}{\mathbf{F}}
\newcommand{\indep}{\perp\!\!\!\perp}
\newcommand{\ind}{\mathbbm{1}}

\let\eps\varepsilon
\let\ph\varphi
\let\Om\Omega
\let\om\omega

\newcommand{\bi}{\operatorname{bi}}
\newcommand{\bilim}{\bi\text{-}\lim}

\begin{document}
\vspace{1.5em}
\begin{center}
\textbf{\MakeUppercase{Approximation of Semigroups of Random Processes on a Hilbert Space}}\\[1em]
\textsc{A.~V.~Utkin}\footnote{E-mail: utkin.av@phystech.edu}\\
\textit{Steklov Mathematical Institute of Russian Academy of Sciences,\\
8 Gubkina St., Moscow 119333, Russia}\\[1.5em]

\section*{Abstract}
{\small
This paper develops Chernoff-type approximation methods for strongly continuous and bi-continuous semigroups of operators arising in the stochastic dynamics of quantum systems. The main focus is on applying Chernoff iterations to Markov semigroups of multiplicative random processes on separable infinite-dimensional Hilbert spaces. The justification of the passage from discrete quantum measurements to continuous ones depends essentially on the choice of topology and function spaces. We propose conditions for convergence to bi-continuous semigroups on subspaces that weaken the classical requirements of the Chernoff theorem. The existence of the limiting bi-continuous semigroup is established
in each of the three cases: commutation of the approximating operators, their asymptotic commutation, and uniform convergence of difference quotients over a family of approximating operators. For diffusive models with a generator having bounded coefficients, the derivative of the approximating function is computed explicitly on the subspace of test functions with rapidly decreasing second derivative; the limitations of applicability of this result are discussed. Applications to generalized models of continuous imprecise quantum measurement with unbounded observables, as well as to Poisson and diffusion processes, are outlined as directions for further research.}
\end{center}

\noindent\textbf{Keywords:} bi-continuous semigroups, Chernoff theorem, Markov operators, stochastic multiplicative processes, quantum measurements.
\vspace{0.5cm}

\section{Introduction}\label{sec:intro}
This work is motivated by the study of stochastic processes on infinite-dimensional
spaces arising in quantum measurement theory and in the theory of open quantum systems. The key problem is to justify the passage from discrete schemes of repeated measurements to continuous-in-time processes.

One of the basic models underlying this passage (the Smolyanov--Truman model,
\cite{SmTr}, \cite{holevo}) is the scheme of repeated imprecise measurements of an
observable, i.e., of a self-adjoint operator $A$. In this scheme, the stochastic
evolution of quantum states is given by the powers of the quantum instrument
\begin{equation}\label{eq_intro:q_instr}
\Mbf(B)[\rho]=\int\limits_B \sqrt{p(A-xI)}\rho\sqrt{p(A-xI)}\ dx,\quad B\in\mathscr{B}(\R),\ \rho\in\mathfrak{S}(\Hilb),
\end{equation}
where $p(x)dx$ is a probability measure and $\mathfrak{S}(\Hilb)$ is the set of positive operators of trace one on the Hilbert space $\Hilb$ (the set of quantum states).

This construction occupies an important place in the theory for two reasons. First, it admits a transparent probabilistic interpretation and the quantum instrument has an explicit form, which makes it a convenient starting point for constructing Chernoff iterations. Second, as the time step is refined ($t\to 0$), the iterations of such an instrument generate continuous-in-time Markov semigroups. The subsequent analysis will also cover more general schemes not reducible to the formula \eqref{eq_intro:q_instr}.

The definition of a continuous measurement process as a limit of discrete ones depends essentially on the chosen type of convergence:
\begin{itemize}
    \item \textbf{Convergence of instrument processes.} In this approach, limiting
    instruments are constructed on the space of trajectories; see \cite{holevo}.
    \item \textbf{Convergence of quantum trajectories}, i.e. random sequences of states $(\rho_1,\rho_2,\ldots)$, in which the passage from the current state to the next one is performed by the quantum instrument in accordance with the measurement outcome obtained; see \cite{attal, pellegrini}.
    \item \textbf{Convergence of Markov operators.} Here one studies the convergence of Markov operators (or of the semigroups formed by them) that arise in the procedure of unraveling the evolution of quantum states; see \cite{unravel1}.
\end{itemize}

In the present work we pursue the third approach --- the approximation of Markov
operators --- building on the ideas of our previous papers \cite{utkin1, utkin2}.

The classical Chernoff theorem \cite{chernoff} provides an effective tool for the
approximation of semigroups by iterations of the form $(\opF_{t/n})^n$. Its application
requires the following two conditions on the derivative $\opF_0'$:
\begin{enumerate}
    \item density of the domain $\Dom(\opF_0')$;
    \item density of the range of $\lambda-\opF_0'$ for some $\lambda>0$.
\end{enumerate}
The second condition is often rather hard to verify. Moreover, in function spaces that are important for applications, Markov semigroups frequently fail to be strongly continuous. Even the diffusion semigroup on $C_b(\R)$, given by
\begin{equation*}\label{eq_intro:diffusion}
(\mathbf{T}_t f)(x)=\int\limits_\R f(x+y)\exp(-y^2/2t)\dfrac{dy}{\sqrt{2\pi t}}.
\end{equation*}
provides a counterexample to strong continuity.

In the present work we develop an alternative approach based on the theory of
bi-continuous semigroups, for which the continuity requirements are weaker.
This makes it possible to work with the spaces $C_b(\Hilb)$, $C_{bws}(\Hilb)$ and
others, in which strong continuity is replaced by bi-continuity. The convergence
analysis is carried out on invariant subspaces (see conditions \ref{def_inv:C.No}),
which substantially simplifies the verification of the hypotheses of the Chernoff
theorem. Instead of the classical requirement of bi-density of the range of
$\lambda-\opF_0'$, we formulate alternative conditions adapted to the specifics of
bi-continuous semigroups on invariant subspaces. The central result is
Theorem \ref{th_inv:main_approx} on approximation on an invariant subspace; its
connection with the convergence of the discrete resolvent is established by
Theorem \ref{th_inv:resolvent_vs_chernoff}, and the case of asymptotic commutation is covered by Theorem \ref{th_a_comm:a_comm_chernoff}.

We note that in \cite{Kolokoltsov2022, Kolokoltsov2026}, the derivation of quantum
filtering equations from sequential measurements is carried out systematically in terms of stochastic equations in spaces of trace-class operators; Chernoff approximation of semigroups is not the main tool there. In the present work, by contrast, the central object of study is the conditions for approximating the semigroups of Markov operators by Chernoff iterations in the most general possible terms, while the limiting filtering equations are not constructed explicitly.

The paper is organized as follows. In Section \ref{sec:design} we fix the notation.
Section \ref{sec:prelim} contains preliminary material: the instrument formalism of
quantum measurements and multiplicative processes (Section \ref{subsec:q_instr}),
bi-admissible spaces and bi-convergence (Section \ref{subsec:bi}), and bi-continuous
semigroups together with the Chernoff theorem for them (Section \ref{subsec:bi_SG}).
Section \ref{sec:main} presents the main results: basic constructions and technical
lemmas (Section \ref{subsec:tech}), approximation on invariant subspaces
(Section \ref{subsec:inv}), and asymptotic commutativity with a converse of the
Chernoff theorem (Section \ref{subsec:a_comm}). Section \ref{sec:examples} deals with the inheritance of invariance (Proposition \ref{prop_inheritance:1}), questions concerning the choice of function spaces and bi-continuity (Example \ref{exa_space:non_cont_SG}, Theorem \ref{th_space:ex}, Proposition \ref{prop_space:bi-cont_UC}), and the diffusion case with an explicit computation of the derivative (Proposition \ref{prop_diff:generator_bounded}). Section \ref{sec:conclusion} summarizes the results and outlines directions for further research.

\section{Designations}\label{sec:design}

This section fixes the basic conventions and notation used throughout the paper.

\paragraph{Hilbert space and operators.}
Let $\Hilb$ be a separable complex Hilbert space. The inner product $(u, v)$ is assumed to be linear in the second argument and antilinear in the first; Dirac notation is not used. The position and momentum operators are denoted by the standard quantum-mechanical symbols $\hat{x}$ and $\hat{p}$. The space of bounded linear operators on $\Hilb$ is denoted by $\mathcal{B}(\Hilb)$, and the space of trace-class operators by $\mathfrak{T}_1(\Hilb)$. The set of quantum states (positive operators of trace one) is denoted by $\Sfr(\Hilb)$.

\paragraph{Typographic conventions.}
To visually distinguish objects related to the Hilbert space $\Hilb$ from those related to function spaces, the following convention is adopted. Operators on $\Hilb$ are typeset in ordinary font (e.g., $A$, $G$). Markov operators, semigroups, generators, and operator-valued functions (OVFs) acting on function spaces are set in boldface (e.g., $\opF_t$, $\mathbf{T}_t$, $\mathbf{A}$); subspaces of function spaces are denoted by a calligraphic font (as $\Dom$).

\paragraph{Function spaces.}
Throughout the paper, an arbitrary Banach space (in particular, any of the function
spaces listed below) is denoted by $X$. The space of continuous linear operators on a Banach space $X$ is denoted by $\mathcal{L}(X)$, in contrast to the space of bounded operators on the Hilbert space $\Hilb$, for which the standard notation
$\mathcal{B}(\Hilb)$ is used.

By $C_b(\Hilb)$ we denote the Banach space of bounded continuous functions
$f:\Hilb\to\Cm$ equipped with the supremum norm. The space $C_{bws}(\Hilb)$ consists of bounded weakly sequentially continuous functions. The spaces of uniformly continuous functions and of functions vanishing at infinity are denoted by $UC_b(\Hilb)$ and $C_0(\Hilb)$, respectively.

\paragraph{Fr\'{e}chet derivatives and tensor powers.}
For functions $f:\Hilb\to\Cm$, the $k$-th order Fr\'{e}chet derivative is denoted by
$\nabla^k f(v)$. Following the approach developed in the works of V.I.~Averbukh,
O.G.~Smolyanov, and S.V.~Fomin, the $k$-th Fr\'{e}chet derivative is identified with
a symmetric $k$-linear functional. By the Riesz representation theorem, such a
functional is uniquely associated with an element of the tensor power $\Hilb^{\otimes k}$. In this context, the notation $(\nabla^k f(v), h^{\otimes k})$ means the application of the $k$-th derivative to the tuple of vectors $(h,\ldots,h)$, which allows one to write the action of differential operators in the infinite-dimensional setting in a uniform and compact way.

\paragraph{Probabilistic structure and random operators.}
A probability space $(\Omega, \mathscr{F}, \Prb)$ is fixed. Random operators $G$ are
regarded as WOT-measurable (i.e., measurable in the weak operator topology) maps
$\Omega \to \mathcal{B}(\Hilb)$, which is equivalent to specifying $\mathcal{B}(\Hilb)$ with a probability measure with respect to which the operators are measurable. The expectation is denoted by $\E$. In writing multiplicative processes, the symbol $\indep$ (e.g., $G_t^{\indep m}$) indicates that the operators in the product are i.i.d. copies. At the same time, in the analysis of continuous limits and of the evolution, it is implicitly assumed for the families $(G_t)_{t\geq 0}$ that the trajectories possess a strong stochastic dependence (consistency) in time.

\paragraph{Quantum measurements.}
Within the instrument formalism (described in detail in Section~\ref{subsec:q_instr}), the action of a quantum instrument on a state $\rho$
with an outcome in a Borel set $B$ is written as $\Mbf(B)[\rho]$.

\paragraph{Families, sequences, and limits.}
Sequences and families of elements are written in parentheses, e.g., $(f_n)_{n\in\N}$. One-parameter families of operators (OVFs) are denoted by subscripts:
$(\opF_t)_{t \geq 0}$. The symbol $\bilim$ denotes the limit in the sense of
bi-convergence, i.e., convergence in a topology $\tau$ on a space $(X,\|\cdot\|,\tau)$ under the additional condition of uniform boundedness in norm; the definition and properties of bi-admissible spaces are deferred to Section~\ref{subsec:bi}.

\section{Preliminaries and Background}\label{sec:prelim}

\subsection{Quantum measurement instruments and multiplicative processes}\label{subsec:q_instr}

Measurements in quantum mechanics are described within the instrument formalism
(see \cite{holevo}). A quantum instrument taking values in a measurable space of
outcomes $(\mathfrak{X}, \mathscr{B})$ is a map
$\Mbf: \mathscr{B}\times \mathfrak{T}_1(\Hilb)\to\mathfrak{T}_1(\Hilb)$ satisfying the following conditions:
\begin{enumerate}
\item for every $B\in\mathscr{B}$, the map $T\mapsto \Mbf(B)[T]$ is linear and
completely positive;
\item for every $T\in\mathfrak{T}_1(\Hilb)$, the map $B\mapsto \Mbf(B)[T]$ is a
$\sigma$-additive (with respect to the WOT) operator-valued measure;
\item $\Mbf(\mathfrak{X})$ is a quantum channel.
\end{enumerate}

The description of a measurement by an instrument $\Mbf$ amounts to two rules:
\begin{enumerate}
\item the probability that the measured value of a system in state $\rho$ falls in
$B\in\mathscr{B}$ is equal to $\Tr \Mbf(B)[\rho]$;
\item the a posteriori state is given by
$\dfrac{\Mbf(B)[\rho]}{\Tr \Mbf(B)[\rho]}$.
\end{enumerate}

If the measured quantity is discrete, the action of the instrument can be represented by a family of completely positive maps $\mathbf{M}_k(\rho) = \sum\limits_m p_{km}G_{km}\rho G_{km}^*$ subject to the normalization condition $\sum\limits_{k,m} p_{km}G_{km}^* G_{km} = I$.

The action of such an instrument induces a stochastic
multiplicative process on the Hilbert space (though not in a unique way):
\begin{equation*}\label{eq_q_instr:psi_k}
\psi_0 \quad \to \quad \psi_1=G_{k_1m_1}\psi_0 \quad \to \quad \psi_2=G_{k_2m_2}\psi_1 \quad \to \quad \ldots
\end{equation*}
with probabilities $1 \quad \to \quad P_1=p_{k_1 m_1} \quad \to \quad P_2=p_{k_1 m_1}P_1 \quad \to \quad \ldots $.

Successive measurements produce a random sequence of outcomes, which corresponds to a quantum trajectory in the state space. In terms of Markov operators, the averaged dynamics of functions $f \in C_b(\Hilb)$ is given by the action
$\opF[G_{(t|s)}]f(v) = \E f(G_{(t|s)}v)$.

Throughout the paper, we assume the normalization condition
\begin{equation}\label{eq_q_instr:q_ch}
\E G_{(t|s)}^*G_{(t|s)}=I.
\end{equation}

In the time-homogeneous case, we arrive at a one-parameter family of Markov operators $\opF_t = \opF[G_t]$: for each $t\geq 0$, $G_t$ is a random bounded operator on $\Hilb$, and
\begin{equation*}\label{eq_q_instr:markov}
(\opF[G_t]f)(v)=\E f(G_t v)=\int f(G_t(\om)v)\ d\Prb(\om).
\end{equation*}

The composition of independent random operators $G_n\ldots G_2G_1$ is again a random
operator, and
\begin{equation*}\label{eq_q_instr:composition}
\opF[G_n\ldots G_1]=\opF[G_1]\ldots\opF[G_n].
\end{equation*}
For a given random operator $G$, there exists a probability measure $P$ on on the Borel $\sigma$-algebra of $\mathcal{B}(\Hilb)$ such that $(\opF[G]f)(v)=\int f(Gv)\ dP(G)$. We write the integral with respect to $P$ as $\E_P$; for a sequence $(G_1,G_2,\ldots)$ of independent $P$-distributed operators defined on a probability space $(\Omega,\mathscr{F},\Prb)$, the product $G_n\ldots G_1$ is denoted by $G^{\indep n}$.

We note that perturbations of quantum dynamical semigroups by covariant completely
positive measures, as well as the problem of reconstruction of states from such an evolution, were studied in \cite{AmosovBaitenovPechen2023}; these constructions also lead to Markov operators of the form \eqref{eq_q_instr:markov}, albeit in a different setting.

\subsection{Bi-admissible spaces and bi-convergence}\label{subsec:bi}

For a precise formulation of the approximation theorems, we introduce the notion of a bi-admissible space (see \cite{diss_bi1}).

\begin{definition}\label{def_bi:bi-adm}
Let $(X,\|\cdot\|)$ be a Banach space, and let $\tau$ be a Hausdorff locally convex
topology on $X$. The triple $(X,\|\cdot\|,\tau)$ is called a bi-admissible space if the following conditions hold:
\begin{enumerate}
\item[\textbf{Bi.1}] the topology $\tau$ is coarser than the norm-topology;
\item[\textbf{Bi.2}] every $\|\cdot\|$-bounded $\tau$-closed set is sequentially $\tau$-complete;
\item[\textbf{Bi.3}] The space $(X,\tau)'$ is norming for $(X,\|\cdot\|)$, i.e. for all $x\in X$
\begin{equation*}\label{eq_bi:norming}
\|x\|=\sup\Bigl\{|\varphi(x)|\ :\ \varphi\in (X,\tau)',\ \|\varphi\|_{X^*}\leq 1\Bigr\}.
\end{equation*}
\end{enumerate}
\end{definition}

The natural type of convergence in such spaces is $\tau$-convergence subject to the
additional condition of $\|\cdot\|$-boundedness. We denote it by
$\bilim\limits_{n\to\infty} x_n = x$.

Several consequences of the definition of a bi-admissible space:
\begin{enumerate}
\item The topology $\tau$ is generated by some family of seminorms $\mathcal{P}_\tau$; moreover, one may assume that $p(x)\leq \|x\|$ for all $p\in\mathcal{P}_\tau$.
\item Condition \textbf{(Bi.2)} is equivalent to the following: every
$\|\cdot\|$-bounded sequence $(x_n)_{n\in\N}$ that is $p$-Cauchy for every $p\in\mathcal{P}_\tau$ converges in $\tau$.
\item The space $(X,\|\cdot\|,\tau_{\|\cdot\|})$ is always bi-admissible; in other words, Banach spaces form a special case of bi-admissible spaces.
\item $(X,\|\cdot\|,\tau_w)$ is bi-admissible whenever $\tau_w$ is the weak topology.
\item If a sequence $(x_n)_{n\in\N}$ bi-converges to $x$, then
$\|x\|\leq \sup\limits_{n\in\N}\|x_n\|$.
\item If a sequence of operators $(\opF_n)_{n\in\N}\subset \mathcal{L}(X)$ satisfies
$C:=\sup\limits_{n\in\N}\|\opF_n\|<\infty$ and the limit $y_x=\bilim\limits_{n\to\infty}\opF_n x$ exists for every $x\in X$, then the map
$x\mapsto y_x$ defines a linear operator $\opF\in\mathcal{L}(X)$ with norm at most $C$.
\item Let a vector-valued function $x(\cdot):[0,T]\to X$ be locally $\|\cdot\|$-bounded $\tau$-continuous ($x(t_0)=\bilim\limits_{t\to t_0}x(t)$ for every $t_0\in[0,T]$). Then the Riemann
integral is well defined by
\begin{equation*}\label{eq_bi:Riemann}
\int\limits_0^T x(t)\ dt=\bilim\limits_{n\to\infty}\sum\limits_{k=1}^n x(kT/n)\ T/n.
\end{equation*}

The Riemann sums $\sum\limits_{k=1}^n x(kT/n)\ T/n$ are bounded in norm by a common
constant and form a $\tau$-Cauchy sequence.
\end{enumerate}

Important examples of bi-admissible spaces in our context are $C_b(\Hilb)$, $C_0(\Hilb)$, $UC_b(\Hilb)$, $C_{bws}(\Hilb)$, and other closed subspaces of $C_b(\Hilb)$ with the supremum norm. The topology $\tau$ may be chosen as
$\tau_b$ (the topology of uniform convergence on bounded sets), $\tau_c$ (the topology of uniform convergence on compact sets), or the norm topology $\tau_{\|\cdot\|}$.

The topology $\tau_b$ is generated by the seminorms
\begin{equation*}\label{eq_bi:tau_b_SN}
\|f\|_n=\sup\limits_{\|v\|\leq n}|f(v)|,\qquad n\in\N,
\end{equation*}
and is therefore metrizable.

\subsection{Bi-continuous semigroups}\label{subsec:bi_SG}

Families of bounded operators are now characterized in terms of their compatibility
with bi-convergence.

\begin{definition}\label{def_bi_SG:equicont}
A family $\mathcal{F}\subset\mathcal{L}(X)$ is bi-equicontinuous if
\begin{enumerate}
\item $\sup\limits_{\opF\in\mathcal{F}}\|\opF\|<\infty$;
\item for every bi-convergent sequence $\bilim\limits_{n\to\infty} x_n=x$ and every
seminorm $p\in \mathcal{P}_\tau$, we have uniform convergence
\begin{equation*}\label{eq_bi_SG:uni}
\lim\limits_{n\to\infty} \sup\limits_{\opF\in\mathcal{F}}p(\opF x_n-\opF  x)=0.
\end{equation*}
\end{enumerate}
\end{definition}

\begin{definition}\label{def_bi_SG:bi-SG}
A semigroup of bounded linear operators $(\mathbf{T}_t)_{t\ge 0}$ on a bi-admissible space $X$ (Definition \ref{def_bi:bi-adm}) is called a bi-continuous semigroup if
\begin{enumerate}
  \item there exist constants $M\geq 1$ and $w\in\R$ such that $\|\mathbf{T}_t\|\leq Me^{wt}$;
  \item the map $t\mapsto \mathbf{T}_t x$ is $\tau$-continuous for every $x\in X$;
  \item for every $T>0$, the family $(\mathbf{T}_t)_{t\in[0,T]}$ is
  bi-equicontinuous.
\end{enumerate}
\end{definition}

Many Markov semigroups of processes on $\Hilb$ are bi-continuous on spaces such as
${(C_b(\Hilb),\|\cdot\|,\tau_b)}$, which makes this framework well suited for the study of stochastic quantum dynamics.

The approximation results below rely on the Chernoff theorem for bi-continuous
semigroups, which we state in its standard form (see, e.g., \cite{diss_bi1, KW, Budde2026}).

\begin{theorem}[Chernoff theorem for bi-continuous semigroups]\label{th_bi_SG:chernoff_bi}
Suppose a family of operators $(\opF_t)_{t\geq 0} \subset \mathcal{L}(X)$ satisfies the following conditions:
\begin{enumerate}
\item $\opF_0 = 1$ is the identity operator;
\item for some $w\in\R$, the family $\bigl((e^{-w t}\opF_t)^m\bigr)_{t>0,m\in\N}$ is bi-equicontinuous;
\item there exists a subspace $\Dom \subset X$ that is bi-dense in $X$ and such
that for every $f\in\Dom$ the bi-limit
\begin{equation*}\label{eq_bi_SG:diff}
\opF_0' f := \bilim\limits_{t\to 0} \dfrac{\opF_t f - f}{t};
\end{equation*}
exists;
\item for some $\lambda>w$, the range $(\lambda-\opF_0')\Dom$ is bi-dense in $X$.
\end{enumerate}
Then there exists a unique bi-continuous semigroup $(\mathbf{T}_t)_{t\ge 0}$ such that for every $f\in X$,
\begin{equation}\label{eq_bi_SG:chernoff}
\mathbf{T}_t f = \bilim\limits_{n\to\infty} (\opF_{t/n})^n f,
\end{equation}
and the convergence is uniform in $t$ on compact intervals in $\R_+$.
\end{theorem}

\section{Main Results}\label{sec:main}

\subsection{Basic constructions and technical lemmas}\label{subsec:tech}

In what follows, we consider bi-admissible spaces with metrizable topologies. The most important example is $(C_b(\Hilb), \|\cdot\|, \tau_b)$ together with its bi-admissible subspaces, where $\tau_b$ is the topology of uniform convergence on bounded sets, generated by the countable family of seminorms
\[
\{\|\cdot\|_{B_n},\ B_n\text{ is the ball of radius }n \text{ centered at the origin}\}.
\]

By contrast, the results of the subsequent sections do not directly apply to the space $(C_b(\Hilb),\|\cdot\|,\tau_c)$ without additional adaptation.

Let $(\opF_t)_{t\geq 0}=\opF[G_t]$ be a one-parameter family of Markov operators, and assume that the normalization condition \eqref{eq_q_instr:q_ch} holds. For a
WOT-measurable random operator $G$, the operator $\opF[G]$ is well defined on
$B_b(\Hilb)$, and the subspaces $C_b(\Hilb)$ and $C_{bws}(\Hilb)$ are
$\opF[G]$-invariant \cite[Proposition 4.2]{utkin1}.

Condition \eqref{eq_q_instr:q_ch} implies uniform stochastic boundedness:
\begin{equation}\label{eq_tech:unif}
  \forall\eps>0,\ r>0,\ T_0>0\ \exists R>0:\quad
  \sup_{\substack{\|v\|\le r,\ t\in[0,T_0]\\ m\in\N:\ mt\le T_0}}
  \Prb\bigl(\|G_t^{\indep m} v\|>R\bigr)<\eps.
\end{equation}

Indeed, by Chebyshev's inequality,
\begin{equation*}\label{eq_tech:chebyshev}
\Prb\Big(\|G_t^{\indep m}v\|>R\Big)\leq \dfrac{\E\bigl(v, (G_t^{\indep m})^*(G_t^{\indep m})v\bigr)}{R^2}=\dfrac{\|v\|^2}{R^2},
\end{equation*}
which implies \eqref{eq_tech:unif}. This property amounts to the uniform tightness of the family of distributions of the random vectors $G_t^{\indep m}v$.

For $\lambda>0$, we introduce the families of approximating operators
\begin{equation*}\label{eq_tech:ch_fam}
\mathcal{F}_\lambda=\Big\{ (e^{-\lambda t}\opF[G_t])^m \quad \Big|\quad t>0,\ m\in\N\Big\}.
\end{equation*}

Condition \eqref{eq_tech:unif} is closely related to the notion of the
compactness containment condition. Let $(X_a)$ be a family of processes on
$[0,+\infty)$ with values in a Banach space $E$. If for every $\eps>0$ and every
$t>0$ there exists a compact set $K_{\eps, t}\subset E$ such that
\begin{equation*}\label{eq_tech:comp_cont}
\inf\limits_{a}\Prb\Big(X_a(t)\in K_{\eps,t}\Big)\geq 1-\eps,
\end{equation*}
then the family is said to satisfy the compactness containment condition.
Verification is often carried out by choosing a Lyapunov function
$V:E\to\R_+$ whose level sets $\{v:\ V(v)\leq c\}$ are compact and for which
\[
\sup\limits_{a, t>0}\E V(X_a(t))<\infty.
\]

In \eqref{eq_tech:unif}, the function $v\to\|v\|^2$ plays precisely the role of such a Lyapunov function.

\begin{proposition}\label{prop_tech:CCC}
Suppose the family of processes $(X_a)$ on a subset of $\Hilb$ equipped with a topology $\tau_0$, where $X_a(t)=G_{t/n}^{\indep n}v$, $a=(v,n)$, satisfies the compactness containment condition. Assume also that the space of $\tau_0$-continuous functions on $\Hilb$ with the supremum norm and the topology $\hat{\tau}_0$ of uniform convergence on $\tau_0$-compact sets, $(C_b(\Hilb,\tau_{\|\cdot\|}),\|\cdot\|,\hat{\tau}_0)$, is bi-admissible.
Then the family $\mathcal{F}_\lambda$ is bi-equicontinuous.
\end{proposition}

\begin{proof}
Fix some $\eps>0$ and a sequence $(f_n)_{n\in\N}$ bi-converging to $f_0$. For pairs
$(m,t)$ with $m\in\N$, $t\in[0,T]$ and $mt>T_\eps=-\dfrac{\ln \eps}{\lambda}$, the corresponding operator $\opF=(e^{-\lambda t}\opF[G_t])^m\in\mathcal{F}_\lambda$  satisfies $\|\opF\|\leq \eps$. Hence,
\begin{equation*}\label{ineq_tech:1}
\Big\|e^{-\lambda mt}\opF[G_t]^m(f_n-f_0)\Big\| \leq 2\eps \sup\limits_{n\in\N}\|f_n\|.
\end{equation*}

If, on the other hand, the pair $(m,t)$ satisfies $mt\in[0,T]$, then for $v\in K$,
where $K\subset\Hilb$ is a $\tau_0$-compact set,
\begin{eqnarray*}\label{ineq_tech:2}
&\left|(e^{-\lambda mt}\opF[G_t]^m(f_n-f_0)(v)\right| \leq\\
&\leq \E \Big[\ind_{\Om_{t,m,v,\eps}}\cdot\left|f_n(G_t^{\indep m}v)-f_0(G_t^{\indep m}v)\right|\Big]+\\
&+\E \Big[(1-\ind_{\Om_{t,m,v,\eps}})\cdot\left|f_n(G_t^{\indep m}v)-f_0(G_t^{\indep m}v)\right|\Big]\leq\\
&\leq  2\eps \sup\limits_{n\in\N}\|f_n\|+\sup\limits_{v\in B_R}|f_n(v)-f_0(v)|,
\end{eqnarray*}
where $\Om_{t,m,v,\eps}=\{\om:\quad G_t^{\indep m}v\notin K_{\eps,t}\}$ is chosen so
that $\Prb(\Om_{t,m,v,\eps})<\eps$ for some $\tau_0$-compact set $K_{\eps,t}$ (which is possible by the compactness containment condition).
Thus, for large $n$, by the uniform convergence of $f_n$ to $f$ on $\tau_0$-compact
sets, the value of the seminorm $\|f\|_K=\sup\limits_{v\in K}|f(v)|$ on the functions $e^{-\lambda mt}\opF[G_t]^m(f_n-f_0)$ becomes arbitrarily small, uniformly in $(m,t)$.
This proves the claim.
\end{proof}

Taking $\tau_0$ to be the weak topology on $\Hilb$, the compactness containment
condition becomes equivalent to \eqref{eq_tech:unif}. If $\tau_0=\tau_{\|\cdot\|}$ is the norm topology on $\Hilb$, one obtains the equicontinuity of $\mathcal{F}_\lambda$ with respect to the topology of uniform convergence on $\|\cdot\|$-compact sets.

Along with the approximating operator-valued function $(\opF_t)_{t\geq 0}$, the
following space associated with the derivative $\opF_0'$ arises naturally:
\begin{equation*}\label{eq_tech:approx_sp}
\Dom(\opF_0')=\Bigl\{\,f\in C_b(\Hilb):\ \exists \opF_0'f:=\bilim_{t\to 0}\frac{\opF_t f-f}{t}\Bigr\}.
\end{equation*}

In \cite{utkin1}, the main approximation result for the commutative case was proved
(Theorem 5.1).

A key step in circumventing the difficulties of verifying the conditions of the
Chernoff theorem was the use of operators forming a discrete analogue of the resolvent of the semigroup. The discrete analogue of the resolvent is obtained by replacing the integral with a sum, while the semigroup operators are replaced by powers of the operators of the approximating OVF. In what follows, we write
\begin{equation*}\label{eq_tech:d_res}
\mathbf{R}^{(N)}_{t,\lambda}=\sum\limits_{m=0}^N te^{-\lambda m t}(\opF_t)^m
\end{equation*}
for $t, \lambda>0$ and $N\in\N\cup\{\infty\}$ (as $N\to\infty$,
$\mathbf{R}^{(N)}_{t,\lambda}f$ converges in norm to $\mathbf{R}^{(\infty)}_{t,\lambda}f$).
These operators also play a key role in Section \ref{subsec:inv}.

\subsection{Approximation on invariant subspaces}\label{subsec:inv}

In the Chernoff theorem, an essential requirement is the bi-density of the range of the operator $\lambda-\opF_0'$; in practice, this requirement is often hard to meet. In this section we show how to overcome this obstacle by a suitable choice of an invariant bi-admissible subspace, adapting the geometry to the requirement.

Choosing, based on practical considerations, a subspace $\Dom\subset\Dom(\opF_0')$, we require that $\opF_t Y \subset Y:=\overline{\Dom}^{\bi}$ for all $t \geq 0$ (which is equivalent to $\opF_t \Dom \subset Y$). This invariance is necessary for the correct restriction of the operators of the family $(\opF_t)_{t\geq 0}$ to the subspace $Y$. If, in addition, one can show that the range $(\lambda-\opF_0')\Dom$ is bi-dense in $Y$, then the structural conditions of the Chernoff theorem are guaranteed to hold.

\begin{definition}\label{def_inv:C.No}
We impose the following conditions on the interplay between the OVF
$(\opF_t)_{t\geq 0}$ and the subspace $\Dom$:
\begin{enumerate}[label=\textbf{C.\arabic*}, ref=\textbf{C.\arabic*}]
\item \label{C:F-comm} \textbf{(commutation condition for the OVF)} For all $t,s>0$,
$[\opF_t,\opF_s]=0$ on $\Dom$.
\item \label{C:core} For $f\in\Dom(\opF_0')\cap Y$ there exists a sequence
$(f_n)_{n\in\N}\subset \Dom$ such that $f=\bilim\limits_{n\to\infty}f_n$ and, for every $t\geq 0$, $\opF_t f\in\Dom(\opF_0')$ and
\begin{equation*}\label{eq_inv:C_core}
\opF_0' \opF_t f=\bilim\limits_{n\to\infty}\opF_0'\opF_t f_n.
\end{equation*}
(In particular, $\Dom$ is a bi-core for $\opF_0'|_Y$.)
\item \label{C:L-comm} \textbf{(commutation condition for the derivative)} For every
$t\geq 0$ and $f\in\Dom$, the vector $\opF_t f$ belongs to $\Dom(\opF_0')$ and
$\opF'_0 \opF_t f=\opF_t \opF_0' f$.
\item \label{C:L-uni} For all $f\in\Dom$, $\lambda>0$ and $\opF\in\mathcal{F}_{\lambda}$, the vector $\opF f$ lies in $\Dom(\opF_0')$, and the bi-convergence of
$\dfrac{\opF_t-1}{t}\opF f$ to $\opF_0'\opF f$ is uniform over
$\opF\in\mathcal{F}_{\lambda}$.
\end{enumerate}
\end{definition}

The conditions introduced above are not independent; we note several relations among
them.

First, condition \ref{C:F-comm} implies \ref{C:L-comm}: the commutation of $\opF_t$ and $\opF_s$ on $\Dom$ yields
\begin{equation*}\label{eq_inv:F_to_L}
\dfrac{\opF_h-1}{h}\opF_t f=\opF_t\frac{\opF_h-1}{h}f\to\opF_t\opF_0'f,
\end{equation*}
whence $\opF_t f\in\Dom(\opF_0')$ and $\opF_0'\opF_t f=\opF_t\opF_0'f$. The converse, however, fails.

Second, in \ref{C:F-comm} the condition $f\in\Dom$ can be replaced with $f\in Y$: by
the bi-density of $\Dom$ in $Y$ and the fact that the operators $\opF_t$ preserve
bi-convergence (and so do their compositions), for $f\in Y$ and $f_n\in\Dom$ with
$f_n\xrightarrow{\bi}f$ we obtain
\begin{equation*}\label{eq_inv:F_comm_to_Y}
[\opF_t,\opF_s]f=\bilim\limits_{n\to\infty}[\opF_t,\opF_s]f_n=0.
\end{equation*}

Finally, if $\opF_0'$ is bi-closed and $\Dom$ is its bi-core (that is, for every $f\in\Dom(\opF_0')\cap Y$ there exists $f_n\in\Dom$ with $f_n\xrightarrow{\bi}f$ and $\opF_0'f_n\xrightarrow{\bi}\opF_0'f$), then \ref{C:L-comm} extends from $\Dom$ to $\Dom(\opF_0')\cap Y$. Indeed, for such $f$ and $(f_n)_n$ we have $\opF_t f_n\xrightarrow{\bi}\opF_t f$ and
\begin{equation*}\label{eq_inv:L_comm_to_Y}
\opF_0'\opF_t f_n=\opF_t\opF_0'f_n\xrightarrow{\bi}\opF_t\opF_0'f,
\end{equation*}
whence, by the bi-closedness of $\opF_0'$, $\opF_t f\in\Dom(\opF_0')$ and
$\opF_0'\opF_t f=\opF_t\opF_0'f$.

\begin{proposition}\label{prop_inv:1}
Suppose one of the following conditions holds:
\begin{enumerate}
\item \ref{C:F-comm};
\item \ref{C:core}+\ref{C:L-comm}.
\end{enumerate}
Then for every $N\in\N$ the image $(\opF_t)^N \Dom$ is contained in $\Dom(\opF_0')$,
and
\[
\opF_0'(\opF_t)^N f=(\opF_t)^N\opF_0'f
\]
for all $f\in\Dom$.
\end{proposition}

\begin{proof}
\begin{enumerate}
\item[\ref{C:F-comm}] For every $N\in\N$,
\begin{equation}\label{eq_inv:eq_1}
\dfrac{\opF_h-1}{h}(\opF_t)^N f=(\opF_t)^N\dfrac{\opF_h-1}{h}f.
\end{equation}
Since $\frac{\opF_h f-f}{h}$ bi-converges to $\opF_0'f$ as $h\to 0$ and $\mathcal{F}_\lambda$ is bi-equicontinuous, the bi-limit on the right-hand side of \eqref{eq_inv:eq_1} exists. Hence $(\opF_t)^N f\in\Dom(\opF_0')$.

\item[\ref{C:core}+\ref{C:L-comm}] Let $t,s\geq 0$ and $f\in\Dom$. By condition
\ref{C:L-comm}, $\opF_t f\in\Dom(\opF_0')$ and $\opF_t \opF_0' f=\opF_0' \opF_t f$.
Then $\opF_t f$ can be bi-approximated by a sequence $(g_n)_{n\in\N}\subset\Dom$ in
such a way that $\opF_0'\opF_s \opF_t f=\bilim\limits_{n\to\infty}\opF_0'\opF_{s}g_n$.
Again, by \ref{C:L-comm}, $\opF_0'\opF_{s}g_n=\opF_{s}\opF_0'g_n$. Using the
bi-equicontinuity, we obtain 
$\bilim\limits_{n\to\infty}\opF_{s}\opF_0'g_n=
\opF_{s}\bilim\limits_{n\to\infty}\opF_0'g_n=
\opF_{s}\opF_0'\opF_t f=\opF_{s}\opF_t\opF_0' f$. 
Thus, $\opF_0'\opF_{s}\opF_t f=\opF_{s}\opF_t\opF_0' f$.

Consequently, for every $N$, $(\opF_t)^N f\in\Dom(\opF_0')$ and
$\opF_0'(\opF_t)^{N}f=(\opF_t)^N\opF_0'f$.
\end{enumerate}
\end{proof}

The result of Proposition \ref{prop_inv:1} also applies to operators of the form
$\mathbf{R}^{(N)}_{t,\lambda}$, which are finite linear combinations of powers of
$\opF_t$.

\begin{proposition}\label{prop_inv:2}
Suppose one of the following alternatives holds:
\begin{enumerate}
\item \ref{C:F-comm};
\item \ref{C:core}+\ref{C:L-comm};
\item \ref{C:L-uni}.
\end{enumerate}
Then for every $\lambda>0$ the subspace $(\lambda-\opF_0')(\Dom(\opF_0')\cap Y)$ is
bi-dense in $Y$.
\end{proposition}

\begin{proof}
\begin{enumerate}
\item[\ref{C:F-comm}]
For $f\in \Dom$, $t,\lambda>0$ and $N\in\N$ we have the identity
\begin{equation}\label{eq_inv:discr}
\mathbf{R}^{(N)}_{t,\lambda}\Bigg[\dfrac{e^{\lambda t}-1}{t}-\dfrac{\opF_t-1}{t}\Bigg]f=e^{\lambda t}f-e^{-\lambda Nt}(\opF_t)^{N+1} f.
\end{equation}

On the left-hand side of \eqref{eq_inv:discr}, the expression in brackets bi-converges to $(\lambda-\opF_0')f$ as $t\to 0$. Note that the operators $\mathbf{R}^{(N)}_{t,\lambda}$ are finite linear combinations of elements of $\mathcal{F}_{\lambda}$; therefore the family $(\mathbf{R}^{(N)}_{t,\lambda})_{t>0, N\in\N}$ is also bi-equicontinuous.

Consequently, for every seminorm $p\in\mathcal{P}_\tau$ and every $\eps>0$,
\begin{equation*}\label{ineq_inv:1}
p\left(\mathbf{R}^{(N)}_{t,\lambda}\Bigg[\dfrac{e^{\lambda t}-1}{t}-\dfrac{\opF_t-1}{t}\Bigg]f-\mathbf{R}^{(N)}_{t,\lambda}(\lambda-\opF_0')f\right)<\eps
\end{equation*}
for sufficiently small $t$, while, in norm, these expressions are uniformly bounded
over $t\in[0,1]$.

Choose a function $N_t\sim 1/t^2$ with values in $\N$. Then
\[
e^{-\lambda N_t t}(\opF_t)^{N_t+1} f \to 0
\]
in norm as $t\to 0$. Passing to the
bi-limit in \eqref{eq_inv:discr}, we obtain
\begin{equation*}\label{eq_inv:eq_2}
f=\bilim\limits_{t\to 0}\mathbf{R}^{(N_t)}_{t,\lambda}(\lambda-\opF_0')f=\bilim\limits_{t\to 0}(\lambda-\opF_0')\mathbf{R}^{(N_t)}_{t,\lambda}f,
\end{equation*}
where, in the last equality, we used the commutation condition \ref{C:F-comm}.

By Proposition \ref{prop_inv:1}, $\mathbf{R}^{(N)}_{t,\lambda}\Dom\subset
\Dom(\opF_0')\cap Y$ for all $N$. Moreover, every $f\in\Dom$ is bi-approximated by
elements of $\bigcup\limits_{N\in\N}(\lambda-\opF_0')\mathbf{R}^{(N)}_{t,\lambda}\Dom$, so that $(\lambda-\opF_0')(\Dom(\opF_0')\cap Y)$ is bi-dense in $Y$, as required.	

\item[\ref{C:core}+\ref{C:L-comm}]
Let $f\in\Dom$.
Choose a function $N_t$, $N_t \xrightarrow[t\to 0]{} \infty$, in such a way that the
remainder $f_{t,\lambda}^{(N_t)}$ in \eqref{eq_inv:discr} tends to zero in norm as
$t\to 0$. Then
\begin{equation*}\label{eq_inv:eq_3}
f=\bilim\limits_{t\to 0}\mathbf{R}_{t,\lambda}^{(N_t)}\Big(\dfrac{e^{\lambda t}-1}{t}-\dfrac{\opF_t-1}{t}\Big)f=\bilim\limits_{t\to 0}\mathbf{R}_{t,\lambda}^{(N_t)}(\lambda-\opF_0')f,
\end{equation*}
where the passage to the limit in the second equality is justified by the
bi-equicontinuity of the family $\left(\mathbf{R}_{t,\lambda}^{(N_t)}\right)_{t>0}$. By Proposition \ref{prop_inv:1}, the commutation gives
$\mathbf{R}_{t,\lambda}^{(N_t)}(\lambda-\opF_0')f=(\lambda-\opF_0')\mathbf{R}_{t,\lambda}^{(N_t)}f$; hence $f$ is a bi-limit of elements of
$\bigcup\limits_{t>0}(\lambda-\opF_0')\mathbf{R}_{t,\lambda}^{(N_t)}\Dom$.

By Proposition \ref{prop_inv:1}, $\mathbf{R}_{t,\lambda}^{(N_t)}\Dom \subset
\Dom(\opF_0')$. Thus $\mathbf{R}_{t,\lambda}^{(N_t)}\Dom \subset
\Dom(\opF_0')\cap Y$, and the set in question is contained in
$(\lambda-\opF_0')(\Dom(\opF_0')\cap Y)$, which proves the assertion, since
$f\in\Dom$, with $\Dom$ bi-dense in $Y$, was arbitrary.

\item[\ref{C:L-uni}]
Fix $\lambda>0$. By condition \ref{C:L-uni}, there exists a constant $C>0$ such that
\begin{equation*}\label{ineq_inv:2}
\sup\limits_{\opF\in\mathcal{F}_{\lambda/2}}\Bigg\|\dfrac{\opF_t-1}{t}\opF f\Bigg\|\leq C.
\end{equation*}

For all $t,s>0$ and $f\in\Dom$,
{\small
\begin{equation}\label{eq_inv:discr2}
\Bigg[\dfrac{e^{\lambda t}-1}{t}\mathbf{R}^{(N)}_{s,\lambda}-\dfrac{\opF_t-1}{t}\mathbf{R}^{(N)}_{s,\lambda}\Bigg]f-e^{\lambda t}f=\Bigg[\dfrac{e^{\lambda t}-1}{t}-\dfrac{\opF_t-1}{t}\Bigg]\Big(\mathbf{R}^{(N)}_{s,\lambda}-\mathbf{R}^{(N)}_{t,\lambda}\Big)f-f_{t,\lambda}^{(N)},
\end{equation}
\par}
where $f_{t,\lambda}^{(N)}=e^{-\lambda N t}(\opF_t)^{N+1} f$ for finite $N$ and
$f_{t,\lambda}^{(\infty)}=0$ (as $N\to \infty$, the remainder $f_{t,\lambda}^{(N)}$
tends to zero in norm).

The remainder of the discrete resolvent series
\begin{equation*}\label{eq_inv:rest}
\mathbf{R}^{(\infty)}_{s,\lambda}-\mathbf{R}^{(N)}_{s,\lambda}=\sum\limits_{m>N} s e^{-\lambda m s}(\opF_s)^m
\end{equation*}
can be written as $\sum\limits_{m>N} s e^{-(\lambda/2) m s}\tilde{\opF}_m$ with
$\tilde{\opF}_m\in\mathcal{F}_{\lambda/2}$. Hence,
\begin{equation*}\label{ineq_inv:3}
\left\|\dfrac{\opF_t-1}{t}\Big(\mathbf{R}^{(\infty)}_{s,\lambda}-\mathbf{R}^{(N)}_{s,\lambda}\Big)f\right\|\leq Cs\sum\limits_{m>N}e^{-(\lambda/2) m s}\xrightarrow[N\to\infty]{}0.
\end{equation*}

Thus,
$\frac{\opF_t-1}{t}(\mathbf{R}^{(\infty)}_{s,\lambda}-\mathbf{R}^{(N)}_{s,\lambda})f\xrightarrow[N\to\infty]{\|\cdot\|}0$
uniformly in $t$ in a neighborhood of zero. On the other hand, for every fixed $N$ we have the bi-convergence
$\frac{\opF_t-1}{t}\mathbf{R}_{s,\lambda}^{(N)}f\xrightarrow[t\to 0]{\bi}\opF_0'\mathbf{R}_{s,\lambda}^{(N)}f$.
Using the uniform bound on the remainder, we conclude that the family
$\frac{\opF_t-1}{t}\mathbf{R}^{(\infty)}_{s,\lambda}f$ converges as $t\to 0$, and
$\mathbf{R}^{(\infty)}_{s,\lambda}f\in\Dom(\opF_0')$.

Consequently, on the left-hand side of \eqref{eq_inv:discr2} with $N=\infty$, the
bi-limit as $t\to 0$ exists and equals $(\lambda-\opF_0')\mathbf{R}^{(\infty)}_{s,\lambda}f-f$.
Passing to the limit on the right-hand side of \eqref{eq_inv:discr2}, we find that this same limit equals
\begin{equation*}\label{eq_inv:bilim_rest}
\bilim\limits_{t\to 0}(\lambda-\opF_0')(\mathbf{R}^{(\infty)}_{s,\lambda}-\mathbf{R}^{(\infty)}_{t,\lambda})f.
\end{equation*}

On the right-hand side, the variables have separated, which allows us to conclude that
$\bilim\limits_{t\to 0}(\lambda-\opF_0')\mathbf{R}^{(\infty)}_{t,\lambda}f$ exists and to pass to the bi-limit as $s\to 0$ on the right-hand side of \eqref{eq_inv:discr2}:
\begin{equation*}\label{eq_inv:bilim_rest2}
\bilim\limits_{s\to 0}(\lambda-\opF_0')\mathbf{R}^{(\infty)}_{s,\lambda}f-f = \bilim\limits_{s\to 0}\bilim\limits_{t\to 0}(\lambda-\opF_0')(\mathbf{R}^{(\infty)}_{s,\lambda}-\mathbf{R}^{(\infty)}_{t,\lambda})f=0.
\end{equation*}

Thus we have shown that the vector $f$ is a bi-limit of elements of the form
$(\lambda-\opF_0')\mathbf{R}^{(\infty)}_{s,\lambda}f$. This yields the asserted
bi-density:
\begin{equation*}\label{eq_inv:L_uni_bi-dense}
\overline{\bigcup\limits_{s>0}(\lambda-\opF_0')\mathbf{R}^{(\infty)}_{s,\lambda}\Dom}^{\bi}=Y\quad \Rightarrow \quad \overline{(\lambda-\opF_0')(\Dom(\opF_0')\cap Y)}^{\bi}=Y.
\end{equation*}
\end{enumerate}
\end{proof}

With the groundwork for verifying the conditions of the Chernoff theorem in place, we state the main approximation result.

\begin{theorem}\label{th_inv:main_approx}
Suppose the OVF $(\opF_t=\opF[G_t])_{t\geq 0}$ satisfies the conditions of Proposition \ref{prop_inv:2}.
Then on $Y=\overline{\Dom}^{\bi}$ there exists a unique bi-continuous semigroup
$(\mathbf{T}_t)_{t\ge 0}$ approximated by the Chernoff iterations
$(\opF_{t/n}^n)_{n\in\N}$ in the sense of bi-convergence:
\begin{equation*}\label{eq_inv:chernoff}
\mathbf{T}_t f=\bilim\limits_{n\to\infty}(\opF_{t/n})^n f,\quad \forall f\in Y,
\end{equation*}
and the convergence is uniform in $t$ on compact intervals in $\R_+$.
\end{theorem}

\begin{proof}
We verify all the conditions of Theorem \ref{th_bi_SG:chernoff_bi}:
\begin{enumerate}
\item The equality $\opF_0 = 1$ holds by the definition of the OVF.
\item The bi-equicontinuity of the family $\mathcal{F}_\lambda$ for every $\lambda>0$ was established in Proposition \ref{prop_tech:CCC}.
\item The existence of the bi-limit of the derivative $\opF_0' f$ on the subspace
$\Dom$ is guaranteed by the choice of $\Dom$.
\item The bi-density of the range $(\lambda-\opF_0')(\Dom(\opF_0')\cap Y)$ in $Y$ for some $\lambda>0$ was established in Proposition \ref{prop_inv:2}.
\end{enumerate}
Thus all the hypotheses of Theorem \ref{th_bi_SG:chernoff_bi} are satisfied, and the
limiting semigroup $\mathbf{T}_t$ exists and is given by the approximation formula
\eqref{eq_bi_SG:chernoff}.
\end{proof}

We now present a result relating the convergence of the discrete resolvent to the
Chernoff approximants; it will be useful for verifying the conditions in concrete
examples.

\begin{theorem}\label{th_inv:resolvent_vs_chernoff}
Suppose condition \ref{C:L-uni} holds; fix $f\in\Dom$.
\begin{enumerate}
\item If for some subsequence $N_k\to\infty$ and some $\lambda>0$ the bi-limit
$g_\lambda(t)=\bilim\limits_{k\to\infty} \mathbf{R}^{(N_k)}_{t/N_k,\lambda}(\lambda-\opF_0')f$
exists, then the Chernoff iterations $(\opF_{t/N_k})^{N_k}f$ bi-converge as
$k\to\infty$, and
\begin{equation*}\label{eq_inv:res_to_SG}
f(t):=\bilim\limits_{k\to\infty}(\opF_{t/N_k})^{N_k}f=e^{\lambda t}\left(f-g_\lambda(t)\right).
\end{equation*}
\item Conversely, if for $t>0$ the bi-limit $f(t):=\bilim\limits_{k\to\infty}(\opF_{t/N_k})^{N_k}f$ exists, then for every $\lambda>0$ and $g_\lambda=(\lambda-\opF_0')f$ there exists the bi-limit $\bilim\limits_{k\to\infty}\mathbf{R}^{(N_k)}_{t/N_k,\lambda}g_\lambda$.
\end{enumerate}
\end{theorem}

\begin{proof}
For any $f\in\Dom$, identity \eqref{eq_inv:discr} with $s_N=t/N$ gives
\begin{equation}\label{eq_inv:discr3}
e^{-\lambda t}(\opF_{s_N})^{N+1}f = e^{\lambda s_N}f - \mathbf{R}^{(N)}_{s_N,\lambda}\left[\dfrac{e^{\lambda s_N}-1}{s_N}-\dfrac{\opF_{s_N}-1}{s_N}\right] f.
\end{equation}
We rewrite the expression in brackets as $\left[\dots\right]f = g_\lambda + \delta_N$, where
\begin{equation*}\label{eq_inv:discr4}
\delta_N := \left(\dfrac{e^{\lambda s_N}-1}{s_N}-\lambda\right)f - \left(\dfrac{\opF_{s_N}-1}{s_N}-\opF_0'\right)f,\quad g_\lambda:=(\lambda-\opF_0')f.
\end{equation*}

Since
$\mathbf{R}^{(N_k)}_{s_{N_k},\lambda} \delta_{N_k}\xrightarrow[k\to\infty]{\bi}0$, we can pass to the limit in \eqref{eq_inv:discr3} along the subsequence
$N_k$, obtaining
\begin{equation*}\label{eq_inv:eq_4}
e^{-\lambda t}f(t) = f - g_\lambda(t),
\end{equation*}
whence $f(t) = e^{\lambda t}(f - g_\lambda(t))$. The first assertion is proved.

The second assertion is proved similarly: if $g_\lambda=(\lambda-\opF_0')f$, then
identity \eqref{eq_inv:discr3} implies the existence of
$\bilim\mathbf{R}^{(N_k)}_{s_k,\lambda}g_\lambda=:g_\lambda(t)$, since the left-hand
side is expressed in terms of $f(t)=\bilim(\opF_{s_k})^{N_k}f$, which exists by
assumption.
\end{proof}

\begin{remark}\label{rem_inv:1}
If the Chernoff iterations bi-converge to a bi-continuous semigroup $\mathbf{T}_t$ with generator $\mathbf{A}$ and resolvent $\mathbf{R}_\lambda$, then $g_\lambda(t)$ is given by the integral $\int\limits_0^t e^{-\lambda s}\mathbf{T}_s (\lambda-\mathbf{A})f\ ds$.
\end{remark}

\subsection{Asymptotic commutativity and an inversion of the Chernoff
theorem}\label{subsec:a_comm}

In many applications, the condition of exact commutation fails, but the operators may commute asymptotically as $t,s\to 0$.
In the context of approximating Markov semigroups on the function space
$C_{bws}(\Hilb)$, the arguments of \cite{utkin2}
simplify owing to the metrizability of the auxiliary topology $\tau$ on a bi-admissible space (which ensures that bi-convergence is characterized by sequences). The attempt to weaken the commutativity condition leads us to a construction of a subspace containing the domain of the derivative $\opF_0'$, defined, however, not in a fully explicit manner, which underscores the importance of studying the approximation conditions for semigroups on subspaces in the general case.
The results of this section were published in \cite{utkin2}; below we present them in a form consistent with the notation of the present paper.

We introduce the subspace $\mathcal{L}_c$ as the set of vectors $f$ for which
\begin{equation*}\label{eq_a_comm:L_c}
\bilim_{t,s\to 0}\frac{[\opF_t,\opF_s]}{ts}\opF_{t_1}\ldots\opF_{t_n}f = 0
\end{equation*}
uniformly over all tuples $(t_1,\ldots,t_n)$ satisfying
$t_1+\ldots +t_n\leq T$ (for an arbitrary fixed $T>0$) and $\max\limits_{1\leq i\leq n} t_i\leq \max(t,s)$.

We also introduce the subspace $\mathcal{L}_a$ of vectors $f$ for which, for every
$d\in\mathbb{Q}_+$, there exists a strictly increasing sequence $(m_k^{(d)})_k$ of positive integers such that the bi-limit $\opF_{t_0/m_k^{(d)}}^{d m_k^{(d)}}f$ exists as $k\to\infty$; in this notation the exponent $d m_k^{(d)}$ is an integer for all sufficiently large $k$.

In the work of A.Yu.~Neklyudov \cite{neklyudov}, an inversion of the Chernoff theorem was obtained: given a weakened form of approximation of a strongly continuous semigroup on a Banach space by Chernoff iterations (by their subsequences), one can assert that they converge in the strong operator topology (SOT) to a strongly continuous semigroup whose generator extends the operator $\opF_0'$. This fact proved useful for verifying the hypotheses of the Chernoff theorem on a suitable subspace, namely, on the bi-closure of $\Dom(\opF_0')\cap\mathcal{L}_a\cap\mathcal{L}_c$. In \cite{utkin2}, the inversion of the
Chernoff theorem was extended to bi-continuous semigroups and their approximants.
We set
\begin{equation*}\label{eq_a_comm:Dom_m}
\Dom_m:=\Dom(\opF_0')\cap\mathcal{L}_a\cap\mathcal{L}_c,\qquad Y_m:=\overline{\Dom_m}^{\bi}.
\end{equation*}
(The symbol $\Dom_m$ is distinct from $\Dom$ in Section \ref{subsec:inv}, where $\Dom$ denoted an arbitrary subspace; $\Dom_m$ is the natural subspace to which the inversion theorem applies.)

\begin{lemma}[\cite{utkin2}, Lemma 3]\label{lem_a_comm:1}
Let $f\in\Dom(\opF_0')$, and suppose that for some sequence $(n_k)_{k\in\N}$ with
$n_k\xrightarrow[k\to\infty]{}\infty$ the bi-limit
\begin{equation*}\label{eq_a_comm:ch_bilim1}
\bilim\limits_{k\to\infty}(\opF_{t_0/n_k})^{n_k}f=\tilde{f}
\end{equation*}
exists.

Then for every sequence $(\tilde{n}_k)_{k\in\N}$ with $\tilde{n}_k\geq n_k$ and
${\tilde{n}_k/n_k\xrightarrow[k\to\infty]{}1}$,
\begin{equation*}\label{eq_a_comm:ch_bilim2}
\bilim\limits_{k\to\infty}(\opF_{t_0/n_k})^{\tilde{n}_k}f=\tilde{f}.
\end{equation*}
\end{lemma} 

Next, consider the linear space $\mathcal{T}$ spanned by formal basis vectors
parameterized by finitely supported sequences of nonnegative numbers
$\vec{t}=(t_1,t_2,\ldots)$.
For $\mathbf{t}=\sum\limits_{k=1}^K c_k\vec{t}_k\in\mathcal{T}$, $c_k\in\Cm$,
$\vec{t}_1=(t_{1,1},t_{1,2},\ldots,t_{1,r_1}),\ \ldots\
,\vec{t}_K=(t_{K,1},t_{K,2},\ldots,t_{K,r_K})$, we define
\begin{equation*}\label{eq_a_comm:max}
\max(\mathbf{t}):=\max\limits_{k,i}t_{k,i},
\end{equation*}
\begin{equation*}\label{eq_a_comm:Len}
\operatorname{Len}(\mathbf{t}):=\max\Big(\sum\limits_{k=1}^K|c_k|, \sum\limits_{k=1}^K|c_k| \sum\limits_i t_{k,i}\Big). 
\end{equation*}

For each element $\mathbf{t}=\sum\limits_{k=1}^K c_k\vec{t}_k$, we write
$\opF_{\mathbf{t}}=\sum\limits_{k=1}^K c_k\opF_{t_{k,1}}\opF_{t_{k,2}}\ldots\opF_{t_{k,r_k}}$.

For each $g\in X$, we define the subspace $X_g^0$ of vectors $f\in X$ admitting a
representation $f=\bilim\limits_{n\to\infty}\opF_{\mathbf{t}_{n}}g$ with
$\max(\mathbf{t}_n)\xrightarrow[n\to\infty]{} 0$ and 
${\sup\limits_{n\in\N}\operatorname{Len}(\mathbf{t}_n)<\infty}$. 
In general, $X_g^0$ may not be a closed subspace, a bi-closed subspace, or invariant under the operators of the family $(\opF_t)_{t\geq 0}$.

\begin{lemma}[\cite{utkin2}, Lemma 4]\label{lem_a_comm:2}
For every $g\in \Dom_m=\Dom(\opF_0')\cap\mathcal{L}_a\cap\mathcal{L}_c$, the subspace $X_g^0$ is contained in $\Dom_m$.
\end{lemma}

The main approximation theorem in this case then reads as follows:

\begin{theorem}[\cite{utkin2}, Theorem 2 and` Corollary 2]\label{th_a_comm:a_comm_chernoff}
Let $\Dom_m=\Dom(\opF_0')\cap\mathcal{L}_a\cap\mathcal{L}_c$ and
$Y_m=\overline{\Dom_m}^{\bi}$. Suppose the OVF $(\opF_t=\opF[G_t])_{t\in[0,T]}$
satisfies $\opF_0=1$ and the normalization condition \eqref{eq_q_instr:q_ch} (in this setting, the local uniform bi-equicontinuity required for the applicability of \cite[Theorem 2]{utkin2} follows from \eqref{eq_q_instr:q_ch}; see Proposition \ref{prop_tech:CCC}). Then on $Y_m$ there exists a unique bi-continuous semigroup $(\mathbf{T}_t)_{t\geq 0}$ such that for every $f\in Y_m$ and every $d\in\mathbb{Q}_+$,
\begin{equation*}\label{eq_a_comm:chernoff}
\mathbf{T}_{t_0 d}f=\bilim\limits_{n\to\infty}\opF_{t_0 d/n}^{n}f,
\end{equation*}
and the map $d\mapsto\mathbf{T}_{t_0 d}$ extends by bi-continuity to all of $\R_+$.
The generator $\mathbf{G}$ of this semigroup extends the operator $\opF_0'|_{Y_m}$.
\end{theorem}

\section{Examples and Applications}\label{sec:examples}

\subsection{Inheritance of invariance}\label{subsec:inheritance}

Establishing or verifying that a sufficiently large subspace $\Dom$ (typically, a
(bi-)core) is invariant under a fixed operator $\opF$ is one of the central problems in applying approximation theorems (e.g., Chernoff's formula) in semigroup theory. This problem is especially acute in infinite-dimensional spaces.

One approach in the context of a Markov operator of the form $\mathbf{F}[G]$ is based on the following fact.

\begin{proposition}\label{prop_inheritance:1}
Suppose a subspace $\Dom\subset C_{bws}(\Hilb)$ is invariant under the map $f(v)\mapsto f(G v)$ $P$-a.s., where $P$ is a probability measure on
$\mathcal{B}(\Hilb)$. Suppose also that $f(v)\mapsto f(G v)$ is locally separably
valued $P$-a.s., i.e., for every $R>0$ there exists a set
$\Omega_{R}\subset \mathcal{B}(\Hilb)$ with $P(\Omega_{R})=1$ such that
$\{(f\circ G)|_{B_R}\ |\ G\in \Omega_{R}\}\subset C_{bw}(B_R)$ is separable.
Then $\opF[G]\Dom\subset\overline{\Dom}^{\bi}$ (and, consequently,
$\overline{\Dom}^{\bi}$ is invariant under $\opF[G]$).
\end{proposition}

\begin{proof}
Fix $f\in\Dom$. By hypothesis, $f\circ G \in \Dom$ and $\|f\circ G\|\leq \|f\|$
$P$-a.s.
For given $r>0$ and $\eps>0$, choose $R>0$ such that
\begin{equation*}\label{ineq_inheritance:1}
P(\|G\|>R/r)<\eps.
\end{equation*}

Since $\Hilb$ is separable, the norm of a random operator is WOT-measurable; hence
$A_R=\{G\ |\ \|G\|\leq R/r\}\subset\mathcal{B}(\Hilb)$ is measurable.
Let $f_R$ denote the restriction of $f$ to the ball $B_R$. Then
$f_R\in C_{bw}(B_R)$ and $f_R(Gv)=f(Gv)$ for all $G\in A_R$, $v\in B_r$.

We show that the map $G\mapsto f_R\circ G$ is weakly measurable in $C_{bw}(B_r)$. Every functional of $(C_{bw}(B_r))^*$ is given by an integral against some bounded measure $\mu$:
\begin{equation*}\label{eq_inheritance:mu}
\ph_\mu(g)=\int\limits_{B_r} g(v)\ d\mu(v).
\end{equation*}

By the dominated convergence theorem and the uniform continuity of $f_R$ in the weak
topology on the ball $B_R$, the map
$G\mapsto \ph_\mu(f_R\circ G)=\int\limits_{B_r}f_R(G v)\ d\mu(v)$ is WOT-continuous,
hence measurable. Consequently, the map $G\mapsto f_R\circ G$ is weakly measurable.
Together with the $P$-a.s. separably valued property, Pettis's measurability theorem
yields the strong measurability of $f_R\circ G$ in $C_{bw}(B_r)$. Since
$\|f_R\circ G\|\leq \|f\|$, the function $G\mapsto f_R\circ G$ is Bochner integrable on $A_R$ (it is extended by zero outside $A_R$). Set
\begin{equation*}\label{eq_inheritance:I_R}
I_R=\int_{A_R} f_R\circ G\ dP(G).
\end{equation*}

Since the Bochner integral lies in the closed convex hull of the values of the
integrand, $I_R$ can be approximated arbitrarily closely by a linear combination of
elements $\{f_R\circ G,\ G\in A_R\}$ with nonnegative coefficients. Choose
$(p_k,G_k)_{k\in\N}$, $G_k\in A_R$, so that $\|I_R-S_R\|_{B_r}<\eps$, where
\begin{equation*}\label{eq_inheritance:S_R}
S_R=\sum\limits_{k=1}^N p_k\ f_R\circ G_k,\quad p_k\geq 0,\ \sum\limits_{k=1}^N p_k=P(A_R).
\end{equation*}

Since $\Dom$ is invariant $P$-a.s., we may choose the $G_k$ outside a null set so that $(f\circ G_k)\in\Dom$ for all $f\in\Dom$.

Since $G_k\in A_R$, we have $f_R(G_k v)=f(G_k v)$ for all $v\in B_r$; hence
\begin{equation*}\label{eq_inheritance:S_R_to_S}
S_R=\sum\limits_{k=1}^N p_k f(G_k v)=:S\in \Dom.
\end{equation*}

It remains to note that
\[
\left\|\int f(Gv)\ dP(G)-I_R\right\|_{B_r}\leq \eps\|f\|\ \text{ and }\ \left\|\int f(Gv)\ dP(G)-I_R\right\|\leq 2\|f\|.
\]

Thus, the function $(\opF[G]f)(v)=\int f(Gv)\ dP(G)$ is approximated uniformly on
$B_r$ by elements $S\in\Dom$ with $\|S\|\leq \|f\|$. Hence
$\opF[G]f\in\overline{\Dom}^{\bi}$.
To prove the invariance of $\overline{\Dom}^{\bi}$, we use the standard step of
bi-approximating by a sequence from $\Dom$.
\end{proof}

\begin{corollary}\label{cor_inheritance:1}
Suppose that $f\circ G$ is locally separably valued and
$(f\circ G)\in\Dom(\opF_0')$ $P$-a.s. Then the bi-closure of the subspace $\Dom(\opF_0')\cap C_{bws}(\Hilb)$ is invariant under $\opF[G]$.
\end{corollary}

\begin{corollary}\label{cor_inheritance:2}
Let $(\xi_1,\ldots,\xi_n)$ be a random vector in $\R^n$, and let
$G_1,\ldots, G_n\in\mathcal{B}(\Hilb)$ be bounded operators. Then
$\sum\limits_{k=1}^n \xi_k G_k$ takes values, with probability $1$, in a separable
subset of $\mathcal{B}(\Hilb)$. For this class of random operators, the set
$\{G(\om):\ \omega\in\Omega\}$ is contained in a finite-dimensional (hence separable) subspace of $\mathcal{B}(\Hilb)$, and the map $G\mapsto f\circ G$ is continuous, so that the local separability condition is satisfied automatically.
\end{corollary}

\subsection{On the bi-continuity of certain semigroups and the choice of spaces}\label{subsec:space}

So far, we have not yet illustrated why the space $C_{bws}(\Hilb)$ is the natural setting for analyzing the convergence of Chernoff iterations to a bi-continuous semigroup. We will show that quite natural examples of stochastic quantum dynamics with unbounded generators lead to semigroups that fail to be bi-continuous on $C_b(\Hilb)$ while being continuous on $C_{bws}(\Hilb)$.
This agrees with the probabilistic interpretation: in a stochastically continuous
adjoint multiplicative process, the time evolution of the characteristic functional of a random vector should be continuous in the sense of pointwise convergence, which is closely related to uniform convergence on bounded subsets. At the same time, it is known that characteristic functionals are weakly sequentially continuous functions on $\Hilb$. Thus, the requirement of bi-continuity of the semigroup of the adjoint process on $C_{bws}(\Hilb)$ arises naturally.

We now present an example of a semigroup on $C_b(\Hilb)$ that is not bi-continuous in the topology of uniform convergence on bounded sets.

\begin{example}\label{exa_space:non_cont_SG}
Let $\hat{x}$ be the position operator in $\mathbb{L}_2(\R)$. Then the Markov
semigroup $\mathbf{T}_t=\opF[G_t]$, where
\begin{equation*}\label{eq_space:non_cont_SG}
G_t=\exp(\eta\sqrt{t} \hat{x}-t\hat{x}^2/2),\quad \eta\sim\mathcal{N}(0,1/2),
\end{equation*}
is not a bi-continuous semigroup on $(C_b(\Hilb),\|\cdot\|,\tau_b)$.
\end{example}

\begin{proof}
We exhibit a counterexample:
\begin{equation*}\label{eq_space:countr_traj}
f(v)=\begin{cases}
0,& \|v\|\leq 1/\sqrt{2} \\
\|v\|^2-1/2,& 1/\sqrt{2}< \|v\|<\sqrt{2}\\
3/2,& \|v\|\geq \sqrt{2} 
\end{cases},\quad f\in C_b(\Hilb).
\end{equation*}

The orbit of $f$ is not $\tau_b$-continuous. Indeed, consider the sequence
\[
v_n(x)=\frac{\sqrt{2}}{\sqrt[4]{\pi}}e^{-(x-n)^2/2},\quad \|v_n\|^2=2.
\]
By construction, $f(v_n)=3/2$.

For any $t>0$, we compute
\begin{eqnarray*}\label{eq_space:sq}
\|G_t(\om)v_n\|^2=& \dfrac{2}{\sqrt{\pi}}\int \exp(2\eta(\om)\sqrt{t}x-tx^2)e^{-(x-n)^2}\ dx=\\
=&\dfrac{2}{\sqrt{1+t}}\exp\Big(\dfrac{\eta^2(\om)t+2\eta(\om)n\sqrt{t}-n^2 t}{1+t}\Big).
\end{eqnarray*}

Hence, as $n\to\infty$, the quantity $\|G_t(\om)v_n\|$ tends to zero a.s., and therefore
\begin{equation*}\label{eq_space:Prb_lim}
\Prb\left(\|G_t(\om)v_n\|<1/\sqrt{2}\right)\to 1.
\end{equation*}
By the dominated convergence theorem, this yields $|\mathbf{T}_t f(v_n)|\to 0$. This
contradicts the $\tau_b$-continuity of the semigroup.
\end{proof}

The following result provides positive statements on $C_{bws}(\Hilb)$.
\begin{theorem}[\cite{utkin1}, Proposition 4.2 and Theorem 4.1]\label{th_space:ex}
\begin{enumerate}
\item The space $C_{bws}(\Hilb)$ is invariant under operators of the form $\opF[G]$.
\item If for the family $(G_t)_{t\geq 0}$ condition \eqref{eq_tech:unif} holds and, for every $\eps>0$ and $w\in\Hilb$,
\begin{equation*}\label{eq_space:cont_Prb}
\lim\limits_{t\to t_0}\Prb\Big(\|(G_t-G_{t_0})^*w\|>\eps\Big)=0,
\end{equation*}
then for all $f\in C_{bws}(\Hilb)$ one has
$\opF[G_t]f\xrightarrow{\bi}\opF[G_{t_0}]f$.
\end{enumerate}
\end{theorem}

Substantive results can also be obtained on other spaces.

\begin{proposition}\label{prop_space:bi-cont_UC}
\begin{enumerate}
\item The spaces $UC_b(\Hilb)$ and $U_bC_b(\Hilb)$ (the space of bounded functions
that are uniformly continuous on balls) are invariant under operators of the form
$\opF[G]$ for any random operator $G$.
\item If $G$ is a random operator whose inverse is bounded $P$-a.s., then the spaces
$C_0(\Hilb)$, $UC_0(\Hilb)$, and $U_bC_0(\Hilb)$ are invariant under $\opF[G]$.
\item Suppose the OVF $(\opF_t=\opF[G_t])_{t\ge 0}$ satisfies
\begin{equation*}\label{eq_space:cont_Prb2}
\lim_{t\to t_0}\sup_{\|v\|\le 1}\Prb\Big(\|G_t v-G_{t_0}v\|>\delta\Big)=0,\quad\forall\delta>0.
\end{equation*}
Then $t\mapsto\opF_t$ is bi-continuous at $t_0$ on
$(U_bC_b(\Hilb),\|\cdot\|,\tau_b)$.
\item If, in addition, $G_t$ has a bounded inverse a.s. for each $t$, then
$t\mapsto\opF_t$ is strongly continuous at $t_0$ on $(U_bC_0(\Hilb),\|\cdot\|)$.
\end{enumerate}
\end{proposition}

\begin{proof}
We first comment on the case of invertible random operators.
The operator $G^{-1}$ is a random operator in the sense of weak measurability of the
map $G^{-1}\colon\Omega\to\mathcal{B}(\Hilb)$; see \cite{hans1} and
\cite[Theorem 9]{hans2}. Consequently, for every $\eps>0$ there exist a measurable set $\Omega_\eps\subset\Omega$ and a constant $q_\eps>0$ such that
$\Prb(\Omega_\eps)>1-\eps$ and for all $\om\in\Omega_\eps$
\begin{equation*}\label{ineq_space:1}
\|G^{-1}(\om)\|\le q_\eps^{-1},\quad \text{ i.e., }\quad \|G(\om)v\|>q_\eps\|v\|\ \forall v\in\Hilb.
\end{equation*}
This uniform lower boundedness property will be key for items 2) and 4).

\begin{enumerate}
\item We prove that $\opF[G]f\in U_bC_b(\Hilb)$ for a random bounded operator $G$ and $f\in U_bC_b(\Hilb)$. Boundedness is obvious.
Let us verify local uniform continuity. Fix $\eps>0$; choose $R>0$ such that
$\Prb(\|G\|>R)<\eps$. By the uniform continuity of $f$ on $B_{R^2}$, there exists
$\delta>0$ such that $|f(u_1)-f(u_2)|<\eps$ whenever $u_1,u_2\in B_{R^2}$ and
$\|u_1-u_2\|<\delta$. Then for $v_1,v_2\in B_R$ with $\|v_1-v_2\|<\delta/R$,
\begin{equation*}\label{ineq_space:2}
|(\opF[G]f)(v_1)-(\opF[G]f)(v_2)|\le \E|f(Gv_1)-f(Gv_2)|.
\end{equation*}
On the set $\{\|G\|\le R\}$ we have $\|Gv_1-Gv_2\|<\delta$ and $Gv_i\in B_{R^2}$, so
$|f(Gv_1)-f(Gv_2)|<\eps$; on the complement (which has probability $<\eps$) the
difference is $\le 2\|f\|$. Hence,
\begin{equation*}\label{ineq_space:3}
|(\opF[G]f)(v_1)-(\opF[G]f)(v_2)|\le \eps(1+2\|f\|).
\end{equation*}
That is, $\opF[G]f$ is uniformly continuous on $B_R$. Since $R$ was arbitrary large, we obtain $\opF[G]f\in U_bC_b(\Hilb)$.

The invariance of $UC_b(\Hilb)$ is proved similarly.

\item We prove the invariance of $C_0(\Hilb)$ assuming that $G$ is invertible a.s.
Suppose $|f(v)|<\eps$ for $\|v\|>R$. As above, for $\|v\|>R /q_\eps$ and
$\om\in\Omega_\eps$ we have $\|G(\om)v\|>R$, so $|f(G(\om)v)|<\eps$. Therefore,
\begin{equation*}\label{ineq_space:4}
|\E f(G v)|\leq \E\Big[\ind_{\Omega_\eps}|f(Gv)|\Big]+\E\Big[(1-\ind_{\Omega_\eps})|f(Gv)|\Big]\leq \eps+\eps\|f\|=\eps(1+\|f\|),
\end{equation*}
where the first inequality uses $|f(G(\om)v)|<\eps$ on $\Omega_\eps$, and the second
uses $\Prb(\Omega_\eps)>1-\eps$.
Consequently, $\lim\limits_{\|v\|\to\infty}|(\opF[G]f)(v)|=0$, i.e.,
$\opF[G]f\in C_0(\Hilb)$.

Since $C_0(\Hilb)$, $UC_b(\Hilb)$, and $U_bC_b(\Hilb)$ are invariant under $\opF[G]$
under the assumptions made, the spaces $UC_0(\Hilb)$ and $U_bC_0(\Hilb)$ are also
invariant, as intersections of invariant subspaces.

\item Fix $f\in U_bC_b(\Hilb)$, $R>0$, and $\eps>0$. Choose $R_0>0$ such that
$\Prb(\|G_{t_0}\|>R_0)<\eps$. The function $f$ is uniformly continuous on
$B_{RR_0+1}$; find $\delta>0$ (assume $\delta<1$) such that $|f(v_1)-f(v_2)|<\eps$
whenever $v_1,v_2\in B_{RR_0+1}$ and $\|v_1-v_2\|<\delta$. In
\eqref{eq_space:cont_Prb2} with threshold $\delta/R$, there exists a neighborhood
$\mathcal U$ of $t_0$ such that for $t\in\mathcal U$ and $\|u\|\le 1$,
\[
\Prb(\|G_tu-G_{t_0}u\|>\delta/R)<\eps;
\]
for $v\in B_R$ this gives
$\Prb(\|G_tv-G_{t_0}v\|>\delta)<\eps$.

For $t\in\mathcal U$ and $v\in B_R$, we have
\[
\Prb(\{\|G_{t_0}\|\le R_0\}\cap\{\|G_tv-G_{t_0}v\|<\delta\})>1-2\eps.
\] 
On this event, $\|G_{t_0}v\|, \|G_tv\|\le RR_0+1$, i.e., both arguments lie in $B_{RR_0+1}$ and their difference has norm $<\delta$; hence $|f(G_tv)-f(G_{t_0}v)|<\eps$.
Therefore,
\begin{equation*}\label{ineq_space:5}
|(\opF_tf)(v)-(\opF_{t_0}f)(v)|\le \eps+4\|f\|\eps
\end{equation*}
for $t\in\mathcal U$ and $v\in B_R$. Consequently,
$\sup\limits_{\|v\|\le R}|(\opF_tf)(v)-(\opF_{t_0}f)(v)|\to 0$ as $t\to t_0$ for every $R>0$, which is precisely $\tau_b$-convergence. Thus, $(\opF_t)_{t\geq 0}$ is
bi-continuous at $t=t_0$.

\item Fix $f\in U_bC_0(\Hilb)$ and $\eps>0$. Choose $R_0$ such that
$\Prb(\|G_{t_0}\|>R_0)<\eps$. Since $f$ vanishes at infinity, there exists $R_1>0$
such that $|f(u)|<\eps$ for $\|u\|>R_1$. Since $G_{t_0}$ is invertible, we can find a measurable set $\Omega_\eps^{t_0}$ with $P_{t_0}(\Omega_\eps^{t_0})>1-\eps$ and a
constant $q_\eps^{t_0}>0$ such that, on $\Omega_\eps^{t_0}$,
$\|G_{t_0}v\|\ge q_\eps^{t_0}\|v\|$ for all $v$. Set $R_*:=2R_1/q_\eps^{t_0}$.

There exists a neighborhood $\mathcal U_0$ of $t_0$ on which, for $\|u\|\le 1$,
$\Prb(\|G_tu-G_{t_0}u\|>q_\eps^{t_0}/2)<\eps$; for $v\ne0$ this gives
\[
\Prb(\|G_tv-G_{t_0}v\|>q_\eps^{t_0}\|v\|/2)<\eps.
\]

The event
$\Omega_\eps^{t_0}\cap\{\|G_{t_0}\|\le R_0\}\cap\{\|G_tv-G_{t_0}v\|\le q_\eps^{t_0}\|v\|/2\}$
has probability $\ge 1-3\eps$. On it, if $\|v\|>R_*$, then
$\|G_{t_0}v\|\ge q_\eps^{t_0}\|v\|>2R_1$, so $|f(G_{t_0}v)|<\eps$; and
\begin{equation*}\label{ineq_space:6}
\|G_tv\|\ge\|G_{t_0}v\|-\|G_tv-G_{t_0}v\|\ge q_\eps^{t_0}\|v\|/2>R_1,
\end{equation*}
whence $|f(G_tv)|<\eps$. Therefore, in this case,
$|f(G_tv)-f(G_{t_0}v)|<2\eps$. On the complement of the event (whose probability is
$<3\eps$), we have $|f(G_tv)-f(G_{t_0}v)|\le 2\|f\|$. Hence, for $t\in\mathcal U_0$ and $\|v\|>R_*$,
\begin{equation*}\label{ineq_space:7}
|(\opF_tf)(v)-(\opF_{t_0}f)(v)|\le 2\eps(1+3\|f\|).
\end{equation*}

We apply the argument of item 3 with the ball $B_{R_*}$: there exists a neighborhood
$\mathcal U_1\subset\mathcal U_0$ of $t_0$ such that for $t\in\mathcal U_1$ and
$\|v\|\le R_*$,
$|(\opF_tf)(v)-(\opF_{t_0}f)(v)|<\eps(1+4\|f\|)$.

Thus, for $t\in\mathcal U_1$ and every $v$,
\begin{equation*}\label{ineq_space:8}
|(\opF_tf)(v)-(\opF_{t_0}f)(v)|\le 2\eps(1+3\|f\|).
\end{equation*}
Hence $\|\opF_tf-\opF_{t_0}f\|\le 2\eps(1+3\|f\|)$. Since $\eps>0$ is arbitrary, we
obtain $\|\opF_tf-\opF_{t_0}f\|\to 0$ as $t\to t_0$.
\end{enumerate}
\end{proof}

\subsection{The diffusion case}\label{subsec:diff}

In this section, we show how to compute the derivative $\opF_0'$ for a wide class of models whose generator is a differential operator with bounded coefficients. This computation is the first step towards verifying conditions \ref{C:F-comm} and \ref{C:L-uni} of Section \ref{subsec:inv}.
The main result of the section is the explicit formula \eqref{eq_diff:generator_formula} for $\opF_0'$ on the class of test functions in $C_b^2(\Hilb)$ with rapidly decaying second derivative. This formula serves as a starting point for verifying the conditions of the approximation theorem:

\begin{itemize}
\item condition \ref{C:L-uni} requires uniform convergence of the difference quotients on the family $\mathcal{F}_{\lambda}$;
\item condition \ref{C:F-comm} is verified directly via the commutation of the random operators $G_t$ and $G_s$; it holds, e.g., if all $G_t$ are functions of mutually commuting self-adjoint operators; in this case, the main Theorem \ref{th_inv:main_approx} applies directly to such models.
\end{itemize}
A complete verification of all the conditions for specific noncommuting models is left for further research.

Let a family of random bounded operators $(G_t)_{t\ge 0}$ be given. We assume the following conditions:
\begin{enumerate}[label=\textbf{(D.\arabic*)}]
\item\label{D:genA1} There exists a bounded operator $A_1\in\mathcal{B}(\Hilb)$ such
that
\begin{equation*}\label{eq_diff:D:genA1}
\left\|\frac1t\E[G_t-I]-A_1\right\|\xrightarrow[t\to 0]{}0.
\end{equation*}
\item\label{D:genA2} There exists a bounded operator $A_2\in\mathcal{B}(\Hilb^{\otimes2})$ such that
\begin{equation*}\label{eq_diff:D:genA2}
\bigl\|\E[(\Delta G_t)^{\otimes2}]/t - A_2^*\bigr\|_{\mathcal{B}(\Hilb^{\otimes2})}\xrightarrow[t\to0]{}0.
\end{equation*}
\item\label{D:bound} $\Prb$-almost surely $\sup\limits_{t\in (0,1)}\|G_t(\om)-I\|/\sqrt{t}\le C$ for some positive constant $C$.
\end{enumerate}

\begin{proposition}\label{prop_diff:generator_bounded}
Suppose conditions \ref{D:genA1}---\ref{D:bound} hold. Then for $f\in C_b^2(\Hilb)$ with $\nabla^2 f$ uniformly continuous on balls and
\begin{equation*}\label{eq_diff:diff2_bound}
\sup_{v\in\Hilb}\|v\|^2\|\nabla^2 f(v)\|<\infty
\end{equation*}
one has
\begin{equation}\label{eq_diff:generator_formula}
(\opF_0'f)(v)=(A_1\nabla f(v),v)+\tfrac12\,(A_2\nabla^2 f(v),v^{\otimes 2}), \quad v\in\Hilb.
\end{equation}
\end{proposition}

\begin{proof}
Set $\Delta G_t=G_t-I$. By Taylor's formula with an integral remainder,
\begin{equation*}\label{eq_diff:tailor}
f(v+\Delta G_tv)=f(v)+(\nabla f(v),\Delta G_tv)+\tfrac12(\nabla^2f(v),(\Delta G_tv)^{\otimes2})+R_t(v),
\end{equation*}
where
\begin{equation*}\label{eq_diff:tailor_rest}
R_t(v)=\int_0^1(1-s)\bigl(\nabla^2f(v+s\Delta G_tv)-\nabla^2f(v),(\Delta G_tv)^{\otimes 2}\bigr)\ ds.
\end{equation*}

Then
\begin{equation}\label{eq_diff:taylor_split}
\frac{\opF_tf(v)-f(v)}{t}=\underbrace{\frac1t\E(\nabla f(v),\Delta G_tv)}_{T_1(t,v)}+\underbrace{\frac1{2t}\E(\nabla^2f(v),(\Delta G_tv)^{\otimes2})}_{T_2(t,v)}+
\underbrace{\frac1t\E R_t(v)}_{R(t,v)}.
\end{equation}

Set $\omega_t(v):=\sup\limits_{s\in[0,1]}\|\nabla^2f(v+s\Delta G_tv)-\nabla^2f(v)\|$. Then $|R_t(v)|\le\|\Delta G_tv\|^2\omega_t(v)$, and
\begin{equation*}\label{ineq_diff:1}
\frac1t\E|R_t(v)|\le\frac1t\E\bigl[\|\Delta G_tv\|^2\omega_t(v)\bigr].
\end{equation*}
By \ref{D:bound}, $\|\Delta G_tv\|^2/t\le C^2\|v\|^2$ a.s. for sufficiently small $t$.
By the uniform continuity of $\nabla^2 f$ on the ball $B_{R(1+C)}$, we have $\sup\limits_{\|v\|\le R}\omega_t(v)\xrightarrow[t\to 0]{}0$ $\Prb$-a.s. (here $\Delta G_tv\to 0$ a.s. by condition \ref{D:bound}); $\omega_t(v)$ is dominated by the integrable function $2\|\nabla^2f\|$. Note that, by the separability of
$\Hilb$ and the uniform continuity of $\nabla^2 f$ on balls, the quantity
$\sup\limits_{\|v\|\le R}\omega_t(v)$ is measurable (it suffices to take the supremum over a countable dense subset).
By the dominated convergence theorem,
\begin{equation*}\label{eq_diff:rest_leb_lim}
\sup_{\|v\|\le R}\frac1t\E|R_t(v)|\le C^2R^2\,\E\left[\sup_{\|v\|\le R}\omega_t(v)\right]\xrightarrow[t\to0]{}0
\end{equation*}

Next, by \ref{D:genA1}, uniformly in $v\in B_R$,
\begin{equation*}\label{eq_diff:lin_lim}
\left(\nabla f(v),\frac1t\E[G_tv-v]\right)\xrightarrow[t\to0]{}(\nabla f(v),A_1v).
\end{equation*}

From \ref{D:genA2} we obtain
\begin{equation*}\label{eq_diff:qua_lim}
\frac1{2t}\bigl(\nabla^2f(v),\E[(\Delta G_tv)^{\otimes2}]\bigr)=\frac12\bigl(\nabla^2f(v),A_2^*v^{\otimes2}\bigr)+o(1)=\tfrac12\bigl(A_2\nabla^2f(v),v^{\otimes2}\bigr)+o(1),
\end{equation*}
where $o(1)$ is uniform in $v\in B_R$ (we used the uniform bound $\|\nabla^2f\|_\infty<\infty$).

Combining the three estimates in \eqref{eq_diff:taylor_split}, we obtain
\emph{uniformly in $v\in B_R$}:
\begin{equation*}\label{eq_diff:general_lim}
\frac{\opF_tf(v)-f(v)}{t}\xrightarrow[t\to0]{}(A_1\nabla f(v),v)+\tfrac12(A_2\nabla^2f(v),v^{\otimes2}).
\end{equation*}

Set $M_2:=\sup_{v}\|v\|^2\|\nabla^2 f(v)\|$. From the inequality
\[
\|\nabla f(r\theta)\|\le\int_r^\infty\|\nabla^2 f(s\theta)\|ds\le\int_r^\infty M_2/s^2\,ds=M_2/r
\]
we obtain $M_1:=\sup_v\|v\|\|\nabla f(v)\|\le M_2<\infty$.
Choose $\delta>0$ such that for $t\in(0,\delta)$ the following hold:
\begin{enumerate}
\item $\|\frac1t\E[G_t-I]-A_1\|\le1$;
\item a.s. $\|\Delta G_t\|\le (C+1)\sqrt t$ and, moreover, $(C+1)\sqrt{\delta}\le1/2$.
\end{enumerate}

For $t\in(0,\delta)$ and every $v\in\Hilb$:
\begin{equation*}\label{eq_diff:T1_maj}
|T_1(t,v)|=\Bigl|\bigl(\nabla f(v),\tfrac1t\E[G_t-I]v\bigr)\Bigr|\le M_1(\|A_1\|+1).
\end{equation*}

From \ref{D:genA2} we obtain $\|\E[(\Delta G_t)^{\otimes2}]/t\|\le\|A_2\|+1$ for
$t\in(0,\delta)$, whence
\begin{equation*}\label{eq_diff:T2_maj}
|T_2(t,v)|\le\tfrac12 M_2(\|A_2\|+1).
\end{equation*}

For $t\in(0,\delta)$ we have $\|v+s\Delta G_tv\|\ge\|v\|/2$ for all $s\in[0,1]$;
therefore, from the condition
$\sup\limits_{u\in\Hilb}\|u\|^2\|\nabla^2 f(u)\|\leq M_2$ 
it follows that
\[
\|v\|^2\|\nabla^2 f(v+s\Delta G_tv)\|\le 4M_2\quad \text{and}\quad \|v\|^2\|\nabla^2 f(v)\|\le M_2,
\]
whence $\|v\|^2\omega_t(v)\le 5M_2$ and
\begin{equation*}\label{eq_diff:rest_maj}
|R(t,v)|\le 5(C+1)^2M_2<\infty.
\end{equation*}

Thus, for $t\in(0,\delta)$ and every $v\in\Hilb$,
\begin{equation*}\label{eq_diff:gen_maj}
\left|\frac{\opF_tf(v)-f(v)}{t}\right|\le K:=M_1(\|A_1\|+1)+\tfrac12 M_2(\|A_2\|+1)+5(C+1)^2M_2<\infty.
\end{equation*}

Together with the $\tau_b$-convergence proved above on each ball $B_R$, this means that the limit $\bilim\limits_{t\to0}\frac{\opF_tf-f}{t}$ exists, so that
$f\in\Dom(\opF_0')$ and formula \eqref{eq_diff:generator_formula} holds.
\end{proof}

\begin{remark}\label{rem_diff:nabla2}
The condition $\sup\limits_v\|v\|^2\|\nabla^2 f(v)\|<\infty$ is essential. For a general $f\in C_b^2(\Hilb)$, the quadratic term $(A_2\nabla^2 f(v),v^{\otimes2})$ grows like $\|v\|^2$, so that the limit $\opF_0'f$ leaves the space $C_b(\Hilb)$, and the $\|\cdot\|$-boundedness of the difference quotients fails. In this case, the limit exists in a weakened sense (pointwise in $v$ and locally in each $\tau_b$-seminorm), but under the standard definition of $\Dom(\opF_0')$, an additional decay of $\nabla^2 f$ is required in order that $f\in\Dom(\opF_0')$.
\end{remark}

\section{Conclusion}\label{sec:conclusion}

The present paper develops an approach to approximating Markov semigroups on
bi-admissible spaces which replaces the hard-to-verify density conditions for the range of $\lambda-\opF_0'$ in the classical Chernoff theorem with commutation conditions (conditions \ref{C:F-comm} and \ref{C:L-comm}), asymptotic commutation (Theorem \ref{th_a_comm:a_comm_chernoff}), and uniform convergence of the difference quotients on the family of approximating operators (condition \ref{C:L-uni}). The main approximation result is stated in Theorem \ref{th_inv:main_approx}; the connection between the convergence of the discrete resolvent and the convergence of Chernoff iterations is established in Theorem \ref{th_inv:resolvent_vs_chernoff}.

For the class of diffusion models with bounded coefficients, the derivative of the approximating function is computed in explicit form (Proposition \ref{prop_diff:generator_bounded}); in the commutative case, this makes it possible to apply the main Theorem \ref{th_inv:main_approx} directly, whereas in the noncommutative case the verification of uniform convergence of the difference quotients is outlined as a direction for further research. 

Further studies will focus on the spaces $\mathcal{L}_a$ and $\mathcal{L}_c$ for processes without strict commutation (see Section \ref{subsec:a_comm}), on extending the explicit computation of
the derivative to models with unbounded coefficients (including continuous imprecise
measurements of position with an unbounded position operator), and on establishing
connections between the Markov semigroups obtained here and Belavkin's stochastic
filtering equations.

We also note that the paper \cite{AmosovSakbaev2025} studies the dynamics of quantum
states, including non-normal states, generated by averaging random operators.
This area contains many open questions. The toolkit developed in the present paper --- approximation of bi-continuous semigroups by Chernoff iterations --- can be applied to some of them.

\section*{Funding}
This study was supported by a grant from the Russian Science Foundation No.~24-11-00145, https://rscf.ru/project/24-11-00145/.

\section*{Acknowledgments}

The author thanks G.\,G.~Amosov and V.\,Zh.~Sakbaev for their support and for
discussions.

In preparing the manuscript, the authors used the DeepSeek model for editing the text, for assistance with the LaTeX code and the bibliography, and for clarifying the conditions of Section \ref{subsec:diff}. The authors checked and edited all the
generated text and bear full responsibility for the content of the publication.

\end{document}